\documentclass[aps,prb,twocolumn,showpacs]{revtex4}
\usepackage{graphicx}
\usepackage{epsfig}
\usepackage{amsmath}
\usepackage{amssymb}
\usepackage{amsfonts}%
\begin{document} 
\title{Effective electron-phonon coupling and polaronic transition in the presence of strong correlation} 
\author{Paolo Barone} 
\affiliation{Dipartimento di Fisica ``E. Amaldi," Universit\`a di Roma Tre, Via della Vasca Navale 84, 00146 Roma, Italy}
\author{Roberto Raimondi}
\affiliation{Dipartimento di Fisica ``E. Amaldi," Universit\`a di Roma Tre, Via della Vasca Navale 84, 00146 Roma, Italy}
\author{Massimo Capone}
\affiliation{INFM-SMC and Istituto dei Sistemi Complessi, Consiglio Nazionale delle Ricerche, Via dei Taurini 19, 00185 Roma, Italy}
\affiliation{Dipartimento di Fisica, Universit\`a ``La Sapienza", P.le Aldo Moro 2, 00185 Roma, Italy}
\author{Claudio Castellani}
\affiliation{Dipartimento di Fisica, Universit\`a ``La Sapienza", P.le Aldo Moro 2, 00185 Roma, Italy}
\begin{abstract} 
We study the Hubbard-Holstein model using slave-boson mean-field and a variational Lang-Firsov
transformation. We identify weak and strong e-ph coupling solutions, whose stability depends 
both on the bare e-ph coupling and on the correlation strength. 
At mean field level the evolution from weak  to strong 
electron-phonon coupling occurs via a first-order polaronic transition if the adiabatic parameter is below a 
critical value.
In the strongly correlated regime and in the adiabatic limit, the region in which the weak-coupling
solution is stable is sizeably enlarged with respect to the weakly correlated system and 
the Mott metal-insulator transition is found to be robust with respect to e-ph interaction. 
\end{abstract} 
\pacs{71.27.+a, 71.10.Fd, 71.30.+h, 71.38.-k} 
 
\date{\today} 
\maketitle 
 
\section{Introduction}

Experimental indications\cite{experiments} of noticeable electron-phonon (e-ph) effects 
in the superconducting cuprates have revived the discussion of the
interplay of e-ph and electron-electron (e-e) interactions.
This issue has obviously a broader relevance, as both interaction mechanisms are important 
also in materials like fullerenes\cite{gunnarson1996} and manganites\cite{millis1996}.
The wealth
of experiments available for the cuprates underscores quite a puzzling scenario where
the fingerprints of e-ph interaction are strikingly different than in weakly correlated materials.
In fact, while some quantities seem to be affected by phonons even more than in ordinary metals,
particularly in the underdoped regime (photoemission spectra \cite{lanzara}, penetration
depth and superfluid density\cite{keller}),
others are almost insensitive to phonons, like resistivity.
Such a situation is ultimately related to the presence of several energy scales in strongly
correlated materials, which makes it compelling  the use of nonperturbative
approaches. 
Several nonperturbative methods have been indeed applied to 
different models with both e-e and e-ph interactions.
Among those, we can count fully numerical approaches such
as various kinds of Quantum Monte Carlo\cite{scalapino,gunnarsson-rosch,misha}, 
exact diagonalization\cite{feschke,io_epj} and
Density-Matrix Renormalization Group\cite{jeckelmann}, the semi-analytical dynamical
mean-field theory (DMFT)\cite{hewson-coreani,capone2004,sangiovanni2005}, and
approximate analytical results based on large-N expansions\cite{kuliczeyher,grilli1994}.
Each of this approaches has its own advantages, and typical range of validity.
An analysis of the recent literature seems to uncover an unclear scenario,
since some calculations conclude an enhanced role of e-ph interaction in
the presence of correlations,\cite{gunnarsson-rosch,misha} and others suggest exactly an opposite situation.\cite{sangiovanni2005}
An important part of those discrepancies can be attributed to the different physical
regimes in which the investigations have been performed
(e.g., antiferromagnetic or paramagnetic phase), but another part
is probably due to the numerical character of the previous analyses
which does not elucidate enough the underlying physical mechanisms.
In the present paper, we try to build a more analytical
understanding for the correlated metal at and near half-filling and compare with 
the results obtained using DMFT in the same regime near the Mott transition.\cite{sangiovanni2005}
DMFT is the quantum version of mean-field theories, that freezes spatial
fluctuations but fully retains local dynamics. The method has been recently 
used to study the effects of e-ph interaction close to 
the Mott metal-insulator transition in the Hubbard-Holstein model. 
\cite{hewson-coreani,sangiovanni2005} 
In particular in Ref. \onlinecite{sangiovanni2005}, it has been
shown that, close to the Mott transition, e-ph interaction manifests in a different 
way in low- and high-energy properties.
While the high-energy Hubbard bands are affected by e-ph coupling, 
the low-energy quasiparticle physics 
is to some extent ``protected" against the phonon effects by e-e correlation.
The main effect of e-ph interaction is indeed the reduction of the
repulsive $U$ leading to an effective interaction
which can be parameterized in a simple way.
Since DMFT still requires numerical calculations 
which sometimes make it difficult to disentangle different
effects, it is instructive to compare approximate analytical
methods with DMFT in order to gain insight into  the physical picture arising from 
the latter method.

To this purpose we present here a mean-field picture of the 
Hubbard-Holstein model, following a recent suggestion by Perroni {\sl et al.}
\cite{cataudella2004}.
From one side we exploit the ability of the slave-boson mean-field approach\cite{kotliar1986}
to capture the main features of the Hubbard model in the strongly correlated regime.
To treat the phonon degrees of freedom we use a variational Lang-Firsov transformation\cite{lang1963},
that has been introduced to  describe the effects of e-ph interaction both in
the weak- and strong-coupling regimes\cite{ciuchi}.
To this end one introduces a ``rotation" parameter 
$f$\cite{cataudella2004}, defined below, which measures the degree of polaron formation.

Slave-boson approaches have been recently performed\cite{baffetto,erik}, in a similar spirit 
of the present paper, to gain insight on the Quantum Monte Carlo results of Ref. \onlinecite{scalapino}, 
proposing that, in a range of correlation values, the small
momentum e-ph vertex can be enhanced by increasing the repulsion 
(yet being smaller than the bare one). The studies of Refs. \onlinecite{baffetto,erik}
suggest that the enhancement of the vertex could be related to a finite-temperature 
signature of a zero-temperature phase separation.
Our approach here follows the spirit of those calculations (namely the use
of simplified approaches to gain insight into more 
involved calculations) but focuses on a slightly different subject, such
as the ``protection" of low-energy degrees of freedom from e-ph effects.

The paper is organized as follows: in Section \ref{method} we introduce the model and derive the general mean-field equations
for the paramagnetic homogeneous phase. In Section \ref{results}, after a brief summary of the most important
results for the half-filling case within the present approach\cite{cataudella2004}, we extend the analysis of the mean-field solutions to the 
strongly correlated small-doping regime and discuss the evolution from weak to strong e-ph coupling, with particular attention to
the way electronic properties are modified. The last Section is devoted to conclusions and acknowledgments.

\section{Mean-field analysis of the Hubbard-Holstein model}\label{method}

The Hamiltonian of the Hubbard-Holstein model reads:
\begin{eqnarray}
H&=&-t\sum_{\langle i,j
\rangle,\sigma}\,c^{\dagger}_{i\sigma}\,c_{j\sigma}\,+U\,\sum_i\,n_{i\uparrow}n_{i\downarrow} \nonumber \\
&+&\omega_0\sum_i\,a^{\dagger}_i\,a_i + \alpha\omega_0\sum_i(a^{\dagger}_i+a_i)\,n_i
\label{ham:model}\end{eqnarray}
where $t$ is a nearest-neighbor hopping term,
$c^{\dagger}_{i\sigma}$ ($c_{i\sigma}$) is the creation (annihilation) operator for an electron with spin $\sigma$
at the site $i$ while $n_i=\sum_\sigma n_{i\sigma}$ is the local density operator. The operator $a_i^{\dagger}$ ($a_i$) 
creates (destroys)
a phonon at the site $i$ with energy $\omega_0$ and $\alpha$ parameterizes the coupling between electrons and
local displacements. In what follows we will introduce the adiabaticity parameter $\gamma=\omega_0/C_dt$ and the 
dimensionless parameter $\lambda=\alpha^2\omega_0/C_d t=\alpha^2\gamma$,  which measures the strength of the
e-ph interaction. Here $C_d$ is a factor depending only on the dimensionality and the topology of the lattice defined 
by the expression of the non-interacting kinetic energy {\it at half filling}, $ |\varepsilon_0|=C_d t$. 

As mentioned above, we treat the e-ph interaction by means of a variational Lang-Firsov transformation
$U=e^S$, with:
\begin{equation}\label{LFtransform}
S=\alpha\sum_i\left[\langle n_i \rangle + f_i\,(n_i-\langle n_i \rangle)\right]\,(a_i - a^\dagger_i).
\end{equation}
$f_i$ are variational parameters which measure the coupling between phonon displacements and density
fluctuations for any value of the adiabaticity parameter. We expect $f_i$ to be equal to one in the antiadiabatic regime,
where the standard Lang-Firsov transformation applies, and equal to zero in the adiabatic regime
where the Born-Oppenheimer approximation holds. Averaging  
$H_{eff}=e^{-S}He^{S}$ on the vacuum state of the transformed phonons 
we obtain a purely fermionic Hubbard-like model with phonon-dependent
renormalizations of the kinetic and correlation terms
and with a rescaling of the chemical potential, all of which are controlled by the variational parameters $f_i$.

The effective electronic model is then solved within the Kotliar-Ruckenstein slave-bosons technique\cite{kotliar1986}. 
For each site we introduce four bosons $e_i,d_i,p_{i\sigma}$, representing the four possible states on site $i$ (zero,
two and one $\sigma$ electron state), and a fermionic operator $\tilde{c}_{i\sigma}$ which is connected to the original electron
operator by the relation $c_{i\sigma}=z_{i\sigma}\tilde{c}_{i\sigma}$, where
\begin{eqnarray}\label{renorfactor}
z_{i\sigma}=\frac{\left(e_i^\dagger p_{i\sigma}  + p_{i\bar{\sigma}}^\dagger d_i
\right)}{\sqrt{1-d_i^\dagger d_i-p^\dagger_{i\sigma}p_{i\sigma}}\sqrt{1-e_i^\dagger
 e_i-p^\dagger_{i\bar{\sigma}} p_{i\bar{\sigma}}}}.
\end{eqnarray}
The equivalence with the original Hilbert space is guaranteed by  the constraints 
\begin{eqnarray}
&&c_{i\sigma}^\dagger c_{i\sigma}=p_{i\sigma}^\dagger p_{i\sigma}+d_i^\dagger d_i  \hspace{1.8truecm} (\forall i,\sigma)\, \nonumber \\
&&1 =\sum_\sigma p_{i\sigma}^\dagger p_{i\sigma}+d_i^\dagger d_i + e_i^\dagger e_i \hspace{1truecm}
(\forall i) \nonumber
\end{eqnarray}
which can be enforced introducing
three Lagrange multipliers $\lambda^{(1)}_i,\lambda^{(2)}_{i\sigma}$.
The mean-field solution at a given value of density $n$ in
the paramagnetic homogeneous phase is obtained by taking the saddle-point value for the 
Bose fields ($\langle e_i\rangle=e_0,\langle d_i\rangle=d_0,\langle p_{i\sigma}\rangle=p_0$)
and assuming translation invariance, so that $f_i=f$ and $\lambda^{(1)}_i=\lambda^{(1)}_0,\lambda^{(2)}_{i\sigma}=\lambda^{(2)}_0$.
Following closely Ref. \onlinecite{lavagna1990}, we minimize
%By minimizing 
the resulting variational energy with respect to the Lagrange multipliers and 
use the constraints to get
%using the standard notations
%\cite{vollhardt1987} $x=e_0+d_0, n=1-\delta$, we find:
\begin{eqnarray}\label{energy}
E_0&=&-|\varepsilon|\,q\,e^{-\alpha^2\,f^2}+d_0^2\,\left[U+2\alpha^2\omega_0(f^2-2f)\right]\nonumber \\
&-& \alpha^2\omega_0(1-\delta)\left[1-\delta(1-f)^2\right]
\end{eqnarray}
where $\varepsilon$
is the kinetic energy per site in the uncorrelated case and $n=1-\delta$; by
introducing the standard notation $x=e_0+d_0$\cite{vollhardt1987}, one can
express the double occupancy as $d_0^2=(x^2-\delta)^2/4x^2$ 
%is the double occupancy, and 
and   $q=z_0^2$, i.e. the reduction of the kinetic energy due to the electronic correlation, as
$q=(2x^2-x^4-\delta^2)/(1-\delta^2)\ $. In the absence of e-ph coupling, when
$\alpha=0$, Eq. (\ref{energy}) reduces to the well-known Gutzwiller energy for the
pure Hubbard model\cite{vollhardt1985}.  On the other hand, taking the limit $f=1$, one recovers the
standard Lang-Firsov result, being the kinetic energy  exponentially renormalized 
with  the electron energy level and the Hubbard term 
shifted respectively by $-\alpha^2\omega_0 $ and $-2\alpha^2\omega_0$.

In order to determine the mean-field solutions we minimize (\ref{energy}) with respect to the remaining variational 
parameters $x^2$ and $f$. One gets:
\begin{eqnarray}
8\,\frac{1-x^2}{1-\delta^2}|\varepsilon|\, e^{-\alpha^2\,f^2} = \left[ U+2\alpha^2\omega_0(f^2-2f)\right]\frac{x^4-\delta^2}{x^4} \label{mfeq:x2}\\
q|\varepsilon|\,f\, e^{-\alpha^2\,f^2} = \omega_0 (1-f)\left[\frac{(x^2-\delta)^2}{2x^2} + \delta(1-\delta) \right].\label{mfeq:f}
\end{eqnarray}
For small doping $\delta$ the kinetic energy can be expanded around the half-filling value $\vert\varepsilon_0\vert$ as 
$\vert\varepsilon\vert=\vert\varepsilon_0\vert(1-a\delta^2)$ (with $a$ depending on the specific shape of the 
uncorrelated band), and the mean-field equation (\ref{mfeq:x2}) can be 
conveniently rewritten as:
\begin{eqnarray}
(1-x^2)\,\frac{x^4}{x^4-\delta^2}\,
=\,\bar{u}\,\frac{1-\delta^2}{1-a\delta^2}\, \label{eq:x2} 
\end{eqnarray}
where $\bar{u}=\left[U/U_c+\lambda (f^2-2f)/4\right]/e^{-\alpha^2\,f^2}$, in 
which $U_c=8\vert\varepsilon_0\vert$ is the Brinkmann-Rice critical value for 
the Mott transition in the absence of phonons. 
Eq. (\ref{eq:x2}) coincides with the result for the Hubbard model 
 once $U/U_c$ is replaced by 
$\bar{u}$ (cfr. Eq. (8) of Ref. \onlinecite{lavagna1990}). This finding is easily interpreted in terms of renormalized
interaction $U_{eff} = U + 2(f^2-2f)\alpha^2\omega_0$ and 
renormalized kinetic energy $\varepsilon_{eff}= \varepsilon e^{-\alpha^2 f^2}$,
showing that the value of the parameter $f$ determines to what extent the electronic
properties are affected by phonons. Furthermore, being Eq. (\ref{mfeq:f}) a 
transcendental equation, it can be useful to separate the exponential from the
algebraic dependence on $f$, obtaining:
\begin{eqnarray}
f = \left[1 +  \frac{2}{\gamma}\,\frac{1-a\delta^2}{1-\delta^2}\,
\frac{x^2\,(2x^2-x^4-\delta^2)}{x^4+\delta^2(1-2x^2)}\,e^{-\alpha^2\,f^2}\right]^{-1}. 
%=\frac{\gamma}{2}\frac{1-\delta^2}{1-a\delta^2}\frac{1-f}{f}
 \label{eq:f}
\end{eqnarray}
This form, as we shall see in the next Section, is more suitable for a graphical analysis.

\section{Results}\label{results}

We are now in the position to discuss the effect of e-ph interaction on 
 the correlation-driven Mott transition at $\delta=0$.
The transition occurs for $\bar{u}=1$, which implies $x^2=0$.
In the small $\gamma$ adiabatic regime if we plug this condition
into Eq. (\ref{eq:f}), we find $f=\gamma /4$\cite{cataudella2004}, while $f=1$ is recovered for
large $\gamma$. 
Therefore, while in the antiadiabatic limit the e-ph coupling strongly 
reduces both the Coulomb repulsion and the kinetic term, in the adiabatic regime, 
the small value of $f$ makes the effect of e-ph smaller on both quantities.
In this regime the exponential factor is closer to one, since the phonons
are less effective in reducing the electron mobility, and, noticeably,
$(U/\varepsilon_0)_{eff} = (U/\varepsilon_0)(1 -\frac{1}{2}\gamma\alpha^2\omega_0/U)$, in remarkable quantitative
agreement with  DMFT results \cite{sangiovanni2005}.
For small $\gamma$ the line marking the Mott transition in the $\lambda$-U diagram 
is then given by $\lambda =(16/\gamma )(U/ U_c -1)$, and reproduces accurately previous diagrams
obtained by DMFT\cite{hewson-coreani}.
The increase of the critical $U$ as a function of e-ph interaction results from the two 
renormalizations of $U_{eff}$ and $\varepsilon_{eff}$.
Because both effects are essential to get the agreement with the DMFT result
at $\delta =0$, we expect them to be  important also at finite $\delta$.

Let us now move away from half-filling, and consider the large $U$ limit.
Here we expect $x^2=\vert\delta\vert/\zeta$ (being $x^2=0$  at half-filling for $u\equiv U/U_c>1$). Expanding (\ref{eq:x2}) to second order in $\delta$
we obtain $\zeta=\sqrt{1-1/\bar{u}}$
that again coincides with the expression for the Hubbard 
model\cite{lavagna1990} with renormalized parameters depending on $f$.
The value of $f$ at small doping is then found by solving  the self-consistent 
equation (derived from Eq. (\ref{eq:f}))
\begin{equation}\label{sol:f}
f=\frac{1}{1+B\frac{2}{\gamma}e^{-\frac{\lambda}{\gamma}f^2}}
\end{equation} 
with $B=2\bar{u}/(2\bar{u}-1)$. In the limit $\bar{u}\gg 1$ 
(which is equivalent to  $u \gg 1$ if $\lambda$ is not too large) the prefactor $B$ goes to $1$,
thus simplifying the analysis of Eq. (\ref{sol:f}). 
 \begin{figure}[ht]
  \centerline{\includegraphics[width=9.0cm]{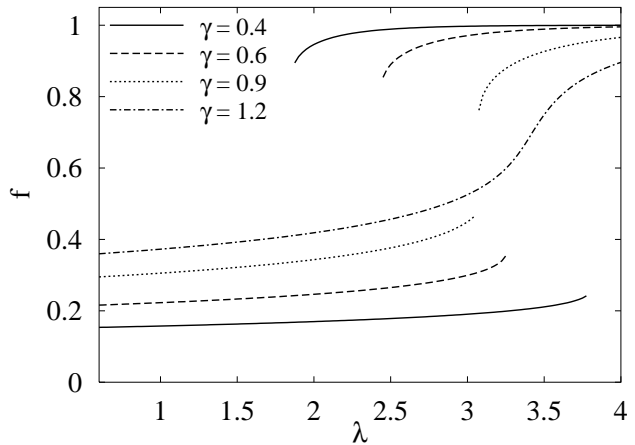}}
  \caption{\label{fig2} Evolution with $\lambda$ of the numerical solution of 
the mean-field equations for $u = 4$ and various values of $\gamma$.  }
\end{figure}

It is useful to compare the behavior of Eq. (\ref{sol:f}) with the 
mean-field theory of the ferromagnetic transition.
We will see that  $f$, $\gamma$ and  $\lambda$
play the roles of the magnetization, temperature and magnetic field, respectively.
The mean-field equation has the form $f=h(f)$ and can be solved graphically 
just like the Curie-Weiss equation. $h(f)$ varies from 0 to 1, so that $f$ 
takes physically  meaningful values.
$h(0)$ is different from zero for any finite value of $\gamma$
and $h(1)<1$. Therefore $h(f)$ intersects the straight line in 
at least one point. 
The function $h(f)$ has an inflection point controlled
by the value of the parameter $\gamma$. 
For small $\gamma$ the inflection point is at small $f$, so that $h(f)$ 
crosses $f$ in three points, while for larger $\gamma$ only one
intersection survives as the inflection point moves toward large $f$.
$\gamma$ clearly plays a role similar to the temperature, with a 
critical value 
$\gamma_c$  which separates the two regimes.
The critical point ($f_c,\gamma_c,\lambda_c$) may be  evaluated  
analytically  by imposing $f_c=h(f_c;\lambda_c,\gamma_c)$, 
$1=\frac{d}{df} h(f_c;\lambda_c,\gamma_c)$, 
and $0=\frac{d^2}{df^2} h(f_c;\lambda_c,\gamma_c)$, which
in the limit $B=1$ give  the  critical values $f_c=\frac{2}{3}$,
$\gamma_c=4 e^{-\frac{3}{2}}\simeq 0.8925$, and $\lambda_c=\frac{27}{8}\gamma_c \simeq 3.0123$.
When three solutions exist, we find two locally stable solutions  
$f_-$ and $f_+$ associated to weaker and stronger effective e-ph
interaction. The two solutions correspond to  negative and positive 
magnetization in the ferromagnetic terminology. 
Notice that the variational nature of our treatment of phonon degrees of 
freedom implies that only the lowest energy state is a valid result of
our approach, even if the equations allow for more solutions.
The parameter $\lambda$ (or more precisely the deviation from $\lambda_c$)
acts as a magnetic field in determining the existence of the two potential
solution and which one is the ground state.
For small (large) $\lambda$ only $f_-$($f_+$) exists, and the energies of 
the two solutions cross in a first order line  $\lambda_c(\gamma)$,
ending in a critical point.
To summarize, in
 the anti-adiabatic limit (large ``temperature") regime, 
we have a single solution smoothly evolving with the ``magnetic field" $\lambda$.
When $\lambda$ is large, lowering the ``temperature'' 
going from the anti-adiabatic to the adiabatic regime
 does not imply a dramatic change in the value of $f$. 
  On the other hand for small $\lambda$ $f$ changes from $f\approx 1$ to 
$f\approx \gamma$ moving from large to small $\gamma$.

\begin{figure}[ht]
\centerline{\includegraphics[width=9.0cm]{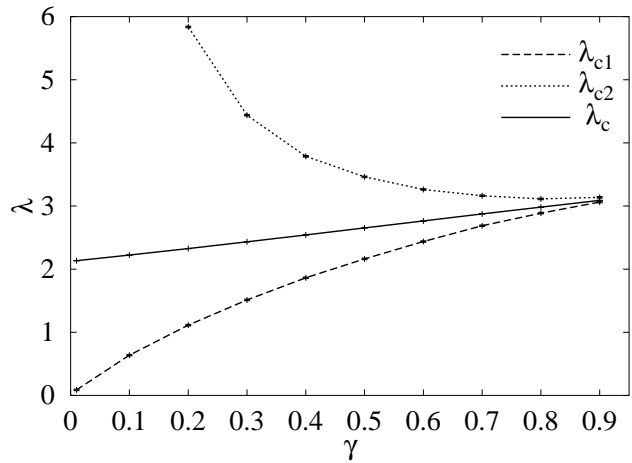}}
\caption{\label{fig3} Phase diagram in the $\gamma-\lambda$
 space at $u = 4$. $\lambda_{c1}$ and 
$\lambda_{c2}$ mark the boundaries of the region in which respectively the $f_+$ and $f_-$ solutions disappear. $\lambda_c(\gamma)$ 
is the transition line where the two solutions have the same energy.  }
\end{figure}

The above ``Curie-Weiss" scenario is recovered also for finite yet large $u = 4$, by
solving numerically the mean-field equations.
In Fig.\ref{fig2} we plot $f$ as a function of $\lambda$ for selected value of 
$\gamma$. As in the $u \gg 1$ limit, the evolution from weak to strong coupling
is smooth for $\gamma > \gamma_c \simeq 0.9$, while for $\gamma < \gamma_c$ the two solutions
with small and large $f$ coexist. Again for large $\lambda$ $f\approx 1$
for any value of the adiabaticity parameter,
while for small values of the e-ph coupling  $f \propto \gamma$ in the adiabatic 
regime and is $f\approx1-1/\gamma$ when $\gamma\gg1$.  
In Fig. \ref{fig3} we report the first-order transition line between 
weak-coupling and strong-coupling solution, as well as the lines $\lambda_{c1}$ and $\lambda_{c2}$ marking
the existence of the two solutions $f_+$ and $f_-$. 
For large $\gamma$ the line of first-order transitions ends in a second-order
end-point, where also $\lambda_{c1}$ and $\lambda_{c2}$ collapse.
The main effect of a finite $u$ compared to the infinite $u$ limit is to 
move the critical point to slightly larger values of
$f,\gamma$ and $\lambda$. 
The strong-coupling solutions always show a full renormalization of $U$, while the kinetic energy is strongly suppressed moving from large to small values of $\gamma$. On the other hand, the weak-coupling solutions are characterized by a
smooth renormalization of the correlation term that goes from its bare value in the adiabatic limit 
to the fully renormalized value $U-2\alpha^2\omega_0$ as $\gamma\rightarrow\infty$; however the 
renormalized kinetic energy $\varepsilon_{eff}$, dominated by the exponential term $\exp (-\lambda f^2 /\gamma )$, 
is always $\approx 1$. 

The phase diagram in Fig. \ref{fig3} is not changed much as function of doping.
This result may be understood in the following way. The phase transition line is obtained by
comparing the values of the energies for the solution for $f_-$ and $f_+$. 
As it can be inferred from Eq. (\ref{energy}),  both the kinetic energy and the polaron energy differ by a quantity of order $\delta$
in the two solutions, so that the critical values of $\lambda$ depend on doping
through subleading corrections in $\delta$ which are quite small at $u = 4$. 
The physics underlying the  phase diagram of  Fig. \ref{fig3}
is that by increasing $\lambda$ a small number $\delta$ of localized holes strongly coupled to phonons
have a lower energy ($\sim- \delta \lambda |\varepsilon_0|$, i.e., the energy of $\delta$ polaronic holes)
than a bad metal that has a reduced kinetic energy, again of order $\delta$ ($\varepsilon_{eff}\sim -\delta |\varepsilon_0|$)
due to the  large effective mass ($m^* \sim 1/\delta$).

We now consider  the fate of the two solutions $f_-$ and $f_+$ as we approach
the Mott transition by letting both $\delta\to 0$ and $U\to U_c$. 
In the adiabatic regime the mean field equations can be solved analytically and 
the polaronic solution only establishes at 
$\lambda_{pol} =4 (U/U_c) (1-\sqrt{1-U_c/U})$. At $U \simeq U_c$ this gives 
$\lambda_{pol} =4$, which is reduced to 
2 in the large $U$ limit. We can translate these
values of the dimensionless coupling in terms of the dimensional phonon-mediated
attraction at the zero-frequency, $V_{ph} = 2\lambda\vert\varepsilon_0\vert$.
We obtain $V_{ph}^{pol} = 8\vert\varepsilon_0\vert = U_c$ close to the Mott transition and 
$V_{ph}^{pol} = 4\vert\varepsilon_0\vert = U_c/2$ in the
large-$U$ limit. 
We therefore find that, for $U$ close to $U_c$, the weak-coupling solution $f_-$ is
always the stable one even for quite large e-ph coupling up to $V_{ph} \simeq U$, thus showing that
the Mott transition is robust with respect to e-ph interaction and polaron formation. \cite{notabipol}
Through the stabilization of the $f_-$ solution the
strong correlations close to the Mott transition also protect the quasiparticles from a strong
mass renormalization due to phonon effects\cite{sangiovanni2005}.
Considering the fluctuations around the mean-field solution
one recovers the high-energy Hubbard bands\cite{raimondi1993}. We expect
that phonon effects will contribute significantly to them\cite{sangiovanni2005}.

As it is well known, the mean-field treatment may be incorrect in  
reproducing the order of a phase transition, and in this sense the polaronic 
transition discussed above may be
turned to a crossover when fluctuations are included, as it happens
in DMFT. On the other hand, the comparison with the DMFT results
shows that the mean-field description captures the basic qualitative features of how the polaron formation is affected 
by the presence of the e-e interaction.

We gain further information on the way the system evolves when the Mott insulator
 is approached by considering the compressibility $\kappa=\partial n/\partial \mu$, or equivalently the 
Landau parameter $F_0^S$, related to $\kappa$ by the relation  
$\kappa=2N^{*}/[1+F_0^s]$, where $N^*$ is the quasiparticle density of states per spin at the Fermi level.
To leading order in $\delta$ we obtain 
\begin{equation}\label{eq:landau}
\frac{F_0^S}{N_0/2\delta}\approx   4|\varepsilon_0|\frac{2\bar{u}-1}{\sqrt{\bar{u}(\bar{u}-1)}}
-16|\varepsilon_0|\alpha^2f(1-f)\frac{\sqrt{\bar{u}(\bar{u}-1)}}{2\bar{u}-1}
\end{equation}
where  $N_0$ is the free DOS (per spin) at the Fermi level.
The second term in Eq. (\ref{eq:landau}) represents the phononic contribution to the Landau parameter dressed by the e-e interaction. The first term 
describes the electronic contribution to $F_0^S$ dressed
by the e-ph correlation; once more we find the same expression of the 
Hubbard model with phonon-renormalized parameters \cite{lavagna1990}.
The attractive phononic contribution to $F_0^S$ increases the compressibility, 
and could eventually lead to a phase separation 
\cite{catabianconi,capone2004}.
This term is indeed only effective for the weak coupling $f_-$ solutions.
However the compressibility of $f_-$ solutions  
diverges only for values of the parameters $\lambda,\gamma$ for which the $f_+$ solution is
energetically favored. This finding is closely reminiscent of the DMFT 
analysis of Ref. \onlinecite{capone2004}, where the insulating behavior is favored 
by the e-ph interaction, but the divergence of the 
compressibility in the metal is prevented by the stabilization of the insulator. 
To verify the accuracy of our approach, we compare our  
compressibility (\ref{eq:landau}) to the predictions of
large-N expansion\cite{grilli1994} and DMFT\cite{capone2004}.
To compare with Ref. \onlinecite{grilli1994}, we take the large-U and small-$\gamma$ 
limit; then (\ref{eq:landau})  acquires a simple random-phase approximation form 
$\kappa(\lambda)=2N^*(0)/[1+(F_0^S)_e + (F_0^S)_{ph}]$, where
$N^*(0)=N_0/2\delta$ is the density of states per spin in the large-$U$ limit in the absence of e-ph interaction
and the phononic
contribution to the symmetric Landau amplitude is simply given by $(F_0^S)_{ph}=-4N^*(0)\alpha^2\omega_0$, in excellent 
agreement with \onlinecite{grilli1994}.  
Relaxing the large-$U$ limit we obtain
\begin{equation}\label{eq:compressibility}
\frac{\kappa(\lambda)}{\kappa(0)}=\frac{1}{1-2\alpha^2\omega_0\beta\kappa(0)}
\end{equation}
where $\kappa(0)$ is the compressibility in the absence of the e-ph interaction 
and $\beta$ is a function of $\gamma$ that approaches unity from below in the adiabatic limit.
Eq. (\ref{eq:compressibility}) is formally identical to Eq. (4) of Ref. \onlinecite{capone2004}, 
where $\beta$ is introduced as a fitting parameter in order to
describe the deviations from the random-phase-like expression for $\kappa(\lambda)$ due to the finite
(as opposed to infinite) value of $U$.

\section{Conclusions}\label{conclusions}

In the present paper we have discussed an analytical approach to the subject of e-ph interaction in
strongly correlated systems, based on slave bosons, treated at the mean-field level, and
a variational Lang-Firsov transformation.
The mean-field description shows that the Mott transition is robust with respect to e-ph interactions,
yielding results in good quantitative agreement with previous DMFT studies\cite{sangiovanni2005}, and supplementing the 
DMFT scenario with some analytical results that make the physical picture more
transparent. 
Furthermore we have shown that the nature of the evolution from weak to strong e-ph coupling is controlled
by the adiabatic parameter $\gamma$. 
In the antiadiabatic regime, i.e., for $\gamma$ larger than a critical value $\gamma_c$, 
the evolution is a smooth crossover, which is replaced, below $\gamma_c$, 
by a first-order transition where the ``rotation" parameter $f$
jumps from a small ($f\simeq  \gamma$, almost no rotation) to a large value ($f \simeq 1$, fully developed polaronic regime).
$f$, despite being simply a variational parameter, measures the effectiveness of the e-ph coupling in affecting the electronic
properties. 

In the adiabatic regime the phonon-mediated attraction which screens the coulomb repulsion is found to 
be reduced proportionally to $\gamma$, in good agreement with DMFT analyses.
In this regime the phonon-induced effective mass enhancement is also reduced 
proportionally to the degree of adiabaticity.
The first-order character of the transition from weak to strong coupling
is most likely a spurious outcome of the mean-field theory.
However the quantitative estimates
of the boundaries between regions with different physical behavior are probably sound: at least they 
provide the correct order of magnitude in those cases where we have accurate
estimates from alternative approaches.
To what extent the present approach can give reliable results far from the 
strongly correlated regime is an interesting
open question but beyond the scope of this paper.

We acknowledge discussions with S. Ciuchi and G. Sangiovanni, and 
financial support from MIUR under Grant No.COFIN2003020239-006.


\begin{thebibliography}{99}
\bibitem{experiments} A.~Lanzara, P.~V.~Bogdanov, X.~J.~Zhou, S.~A.~Kellar, D.~L.~Feng, 
E.~D.~Lu, T.~Yoshida, H.~Eisaki, A.~Fujimori, K.~Kishio, J.~-I.~Shimoyama, T.~Noda, S.~Uchida, Z.~Hussain, and Z.~X.~Shen,
Nature {\bf 412}, 510 (2001); M.~d'Astuto,  P.~K.~Mang, P.~Giura, A.~Shukla, P.~Ghigna, 
A.~Mirone, M.~Braden, M.~Greven, M.~Krisch,  and F.~Sette,
Phys. Rev. Lett. {\bf 88}, 167002 (2002).

\bibitem{gunnarson1996}  O.~Gunnarsson, Rev. Mod. Phys. {\bf 69}, 575 (1997).

\bibitem{millis1996} A.~J.~Millis, R.~Mueller, and B.~I.~Shraiman
Phys. Rev. B {\bf 54}, 5405 (1996).

\bibitem{lanzara}  G.~-H.~Gweon, T.~Sasagawa, S.~Y.~Zhou, J.~Graf, H.~Takagi, D.~-H.~Lee, A.~Lanzara, Nature {\bf 430}, 187 (2004).

\bibitem{keller}
G.~M.~Zhao, M.~B.~Hunt, H.~Keller, K.~A.~M\"uller, Nature (London) {\bf 385}, 236 (1997); 
R.~Khasanov, D.~G.~Eshchenko, H.~Luetkens, E.~Morenzoni, T.~Prokscha, A.~Suter, N.~Garifianov, M.~Mali, J.~Roos, 
K.~Conder, and H.~Keller, Phys. Rev. Lett. {\bf 92}, 057602 (2004).

\bibitem{scalapino} Z.~B.~Huang, W.~Hanke, E.~Arrigoni and D.~J.~Scalapino,
Phys. Rev. B {\bf 68}, 220507(R) (2003).

\bibitem{gunnarsson-rosch} O.~R\"osch and O.~Gunnarsson, Phys. Rev. Lett. {\bf 93}, 237001 (2004).

\bibitem{misha} A.~S.~Mishchenko and N.~Nagaosa, Phys. Rev. Lett. {\bf 93}, 036402 (2004).

\bibitem{feschke} B.~B\"auml, G.~Wellein, and H.~Fehske, Phys. Rev. B {\bf 58}, 3663 (1998).

\bibitem{io_epj} M.~Capone, M.~Grilli, and W.~Stephan, Eur. Phys. J. B {\bf 11}, 551 (1999).

\bibitem{jeckelmann} E.~Jeckelmann, Phys. Rev. B {\bf 57}, 11838 (1998).

\bibitem{hewson-coreani} W.~Koller,~D. Meyer and A.~C.~Hewson, Phys. Rev. B {\bf 70}, 155103 (2004);
G.~S.~Jeon,  T.~-H.~Park, J.~H.~Han, H.~C.~Lee, and H.~-Y.~Choi, Phys. Rev. B {\bf 70}, 125114 (2004).

\bibitem{capone2004} M.~Capone, G.~Sangiovanni, C.~Castellani, C.~Di Castro, and M.~Grilli
Phys. Rev. Lett. {\bf 92},106401 (2004).

\bibitem{sangiovanni2005}  G.~Sangiovanni, M.~Capone, C.~Castellani, M.~Grilli, 
Phys. Rev. Lett. {\bf 94}, 026401 (2005).

\bibitem{kuliczeyher} R.~Zeyher and M.~L.~Kuli\'c, Phys. Rev. B {\bf 53}, 2850 (1996)

\bibitem{grilli1994} M.~Grilli and C.~Castellani, Phys. Rev. B {\bf 50}, 16880 (1994).

\bibitem{cataudella2004} C.~A.~Perroni, V.~Cataudella, G.~De Filippis and 
V.~Marigliano~Ramaglia, Phys. Rev. B {\bf 71}, 113107 (2005).

\bibitem{kotliar1986} G.~Kotliar and A.~E.~Ruckenstein, Phys. Rev. Lett. {\bf 57},1362 (1986).

\bibitem{lang1963} I.~J.~Lang and Yu.~A.~Firsov, Soviet Physics JETP {\bf 16}, 1301 (1963).

\bibitem{ciuchi} D.~Feinberg, S.~Ciuchi, F.~de~Pasquale, Int. J. Mod. Phys. B {\bf 4},1317 (1990).

\bibitem{baffetto}  E.~Cappelluti, B.~Cerruti, and L.~Pietronero,
Phys. Rev. B {\bf 69}, 161101(R) (2004).

\bibitem{erik} E.~Koch and R.~Zeyher, Phys. Rev. B {\bf 70}, 094510 (2004).

\bibitem{lavagna1990} M.~Lavagna, Phys. Rev. B {\bf 41},142 (1991).

\bibitem{vollhardt1987} D.~Vollhardt, P.~Wolfle and P.~W.~Anderson, 
Phys. Rev. B {\bf 35}, 6703 (1987).

\bibitem{vollhardt1985} For a pedagogical introduction see D.~Vollhardt, Rev. Mod. Phys. {\bf 56}, 99 (1984). 

\bibitem{notabipol} Notice that for
$V_{ph} > U$ the attraction exceeds the repulsion and the system is expected to switch to a bipolaronic regime.

\bibitem{raimondi1993}R. Raimondi and C. Castellani, Phys. Rev. B {\bf 48}, R11453 (1993) 


\bibitem{catabianconi} V.~Cataudella, G.~De~Filippis, G.~Iadonisi, A.~Bianconi, and N.~L.~Saini, Int. J. Mod. Phys. B {\bf 14}, 3398 (2000). 



\end{thebibliography}
\end{document}